\documentclass[onecolumn,superscriptaddress,showpacs,preprintnumbers]{revtex4-1}
\usepackage{soul}
\usepackage{amsmath}
\usepackage{amssymb}
\usepackage{graphicx}
\usepackage{morefloats}
\usepackage[usenames,dvipsnames]{xcolor}
\usepackage{blkarray}
\usepackage{verbatim}
\usepackage{hyperref}
\usepackage{subfigure}
\usepackage{pifont}

\usepackage{natbib}
\usepackage{wrapfig}
\usepackage{mathrsfs}
\usepackage{tikz}
\usepackage{soul}
\usepackage{booktabs}
\usetikzlibrary{arrows}

\hypersetup{colorlinks=blue, citecolor=orange, urlcolor=cyan, linkcolor=magenta}

\begin{document}

\title{Extracting significant signal of news consumption from social networks:\\the case of Twitter in Italian political elections}
\author{Carolina Becatti}
\affiliation{IMT School for Advanced Studies, P.zza S. Francesco 19, 55100 Lucca (Italy)}
\author{Guido Caldarelli}
\affiliation{IMT School for Advanced Studies, P.zza S. Francesco 19, 55100 Lucca (Italy)}
\affiliation{Istituto dei Sistemi Complessi (CNR) UoS Universit\`a “Sapienza”, P.le A. Moro 2, 00185 Rome (Italy)}
\affiliation{European Centre for Living Technology, Universit\`a di Venezia “Ca' Foscari”, S. Marco 2940, 30124 Venice (Italy)}
\affiliation{Catchy srl, Talent Garden Poste Italiane Via Giuseppe Andreoli 9, 00195 Rome (Italy)}
\author{Renaud Lambiotte}
\affiliation{Mathematical Institute, University of Oxford, Oxford (UK)}
\author{Fabio Saracco}
\affiliation{IMT School for Advanced Studies, P.zza S. Francesco 19, 55100 Lucca (Italy)}

\begin{abstract}
\vspace{1cm}
According to the Eurobarometer report about EU media use of May 2018, the number of European citizens who consult on-line social networks for accessing information is considerably increasing. In this work we analyze approximately $10^6$ tweets exchanged during the last Italian elections. By using an entropy-based null model discounting the activity of the users, we first identify potential political alliances within the group of verified accounts: if two verified users are retweeted more than expected by the non-verified ones, they are likely to be related. Then, we derive the users' affiliation to a coalition measuring the polarization of unverified accounts. Finally, we study the bipartite directed representation of the tweets and retweets network, in which tweets and users are collected on the two layers. Users with the highest out-degree identify the most popular ones, whereas highest out-degree posts are the most ``viral". We identify significant content spreaders by statistically validating the connections that cannot be explained by users' tweeting activity and posts' virality by using an entropy-based null model as benchmark. The analysis of the directed network of validated retweets reveals signals of the alliances formed after the elections, highlighting commonalities of interests before the event of the national elections.
\end{abstract}

\maketitle

\section{Introduction}
In recent years, the computer revolution we are witnessing made evident that 
the interconnection between individuals in the society formed a different (non regular) lattice on which news and gossip propagate from one person to another one \cite{barabasi2011network}. 
Indeed, the way Europeans approach news and information has drastically changed. According to the Eurobarometer report about EU media use of May 2018~\cite{TNSopinionsocial2018}, printed press is consulted everyday by 28\% of the EU citizens, following  a decreasing trend, while is never consulted by approximately 20\% of them. On the other hand, the daily access to the Internet increased up to the 65\% of the population (starting from a 45\% in 2010) and the increase in the percentage of citizens using everyday on-line social network for accessing information is even more striking, passing from 18\% in 2010 to 42\% in 2017. This movement towards new technologies, beside being different from country to country (for instance, in Belgium the everyday access to the web concerns the 72\% of the population against the 53\% of Italy), is paradoxically in opposition to a general distrust towards new media: only 20\% of the population declared to trust the contents available in on-line social networks, 34\% trusts the Internet in general, while radio (59\%) and TV (51\%) are considered as more reliable.\\
\indent In order to provide a global understanding of the structure and the dynamic of these new media, complex network studies~\cite{Newman2010,caldarelli2010scale-free} tackled different aspects of these phenomena. Different studies analyzed the raising of grass-root democratic protests as Arab springs~\cite{Gonzalez-Bailon2011}, Occupy Wall Street~\cite{Conover2011} or the Spanish ``Indignados"~\cite{Gonzalez-Bailon2013}. Analogously, the dynamics of the electoral campaign online media has a (relatively) long literature, focusing, time by time, on USA  ~\cite{Adamic2005,Diakopoulos2010,DiGrazia2013, Bekafigo2013, Badawy2018, Bovet2018,Bovet2018a}, Australia~\cite{Gibson2006,Bruns2012}, Norway~\cite{Enli2013}, 
Spain~\cite{Borondo2012}, Italy~\cite{Caldarelli2014,DelVicario2017d,Bindi2018}, France~\cite{Ferrara2017} and UK~\cite{Cram2017,DelVicario2017c}.
Generally, the  shift from mediated to disintermediated news consumption has led to a range of documented phenomena: users tend to focus on information reinforcing their opinion (\emph{confirmation bias}~\cite{Quattrociocchi2015,DelVicario2016,DelVicario2016a, Schmidt2018,Schmidt2018a}) and to  
group in clusters of people with similar viewpoints, forming the so called  \emph{echo chambers}~\cite{Bakshy2015,Nikolov2015,DelVicario2016,DelVicario2016a, Schmidt2018,Schmidt2018a}. The different dynamics that the public debate follows on social-network platforms is also remarkable: the time evolution of viral non-verified contents is more persistent than the verified equivalent~\cite{DelVicario2016a} and ``negative" messages spread faster than ``positive" ones, even if the latter reach on average a wider audience~\cite{Ferrara2015}. Moreover, the analysis of time evolution of the activities in social platforms helps to predict the trend of retweets~\cite{Kobayashi2016}, the interactions of a single user with her/his neighbours~\cite{Tabourier2016} and to detect future developments of information campaign at an early stage~\cite{Varol2017} or ``astroturf" campaigns~\cite{Ferrara2017}.\\
\indent Extracting information from on-line social networks is often complicated by their complex, intertwined organization, and their strong heterogeneity. Salient features can be  identified as significant deviations from carefully-constructed null models. An unbiased entropy-based null-model providing  a general framework for the analysis of real-world networks, can be obtained following the information theory derivation of statistical physics~\cite{Jaynes1957,Squartinia,Cimini2018}: starting from an observed network, define an ensemble comprehending all possible networks with the same amount of nodes \footnote{Nodes play the role of the volume in the canonical definition in statistical physics.}, but varying the number of links, passing from an empty network to the fully connected one, via all possible link configurations. Then define a probability distribution over the ensemble; the functional form of this distribution can be derived through the maximization of the entropy under certain constraints~\cite{park2004statistical}, i.e. preserving the average values over the ensemble of some quantities of interest. In order to obtain an estimate for the probability distribution parameters, we maximize the likelihood to observe the real network~\cite{Garlaschelli2008, squartini2011analytical}.
The crucial point of the above construction is the role of the constraints: in order to provide a reliable null-model, constraints should represent important properties of the system under analysis. Depending on the application, they may represent either the total number of links, as in the Erd\"os-R\'enyi null model, or the degree sequence~\cite{park2004statistical,Garlaschelli2008,squartini2011analytical}, or other topological properties~\cite{Squartini2013,Fronczak2013a, Mastrandrea2014,DiGangi2018,Becatti2018,DeJeude2018}.\\ 
\indent In the present work, we analyze $\sim10^6$ tweets exchanged during the last Italian elections. Using entropy-based null-models, we detect non-trivial patterns in the diffusion of viral contents and different diffusion strategies depending on the polarization of the users. We employ a first undirected representation of the network of retweets by distinguishing certified from non-certified users. The former group is mostly composed of celebrities and official accounts, including politicians, newspapers, etc. 
Then we identify groups of verified users by their interaction with the opposite layer, following the recipe of~\cite{Gualdi2016a,Saracco2016}. If two verified users are retweeted more than expected by non-verified ones, they are likely to be related. 
We analyse the  community organisation of the resulting network and measure the polarization of unverified users: as observed in other studies~\cite{Dandekar2012,Flaxman2013,Bakshy2015,Nikolov2015,Quattrociocchi2015,DelVicario2016,DelVicario2016a,Schmidt2018,Schmidt2018a}, people tend to interact just with a single community, strongly polarizing their opinions and we confirm this observation in the Italian elections of 2018.
Finally, we study a bipartite and directed representation of the users' tweets and retweets network, in order to identify significant news consumers. In this context, we detect both the most popular and most significant content spreaders
for the different communities. In the former case, we focus only on the users with the highest out-degree,
in the latter case we statistically validate the (directed) connections that cannot be simply explained by the ``virality" of the tweets and the tweet/retweet activity of the users. This last validation is an extension of the approach of~\cite{Gualdi2016a,Saracco2016} to direct bipartite networks.\\
\indent Our analysis uncovers the various strategies for the electoral campaign, followed by different political alliances and highlight a different participation to the political debate, providing indications about the role of the  users to spread the viral content inside each community. Moreover, by analysing the structure of the directed network of statistically validated retweets, we observe not only the alliances presented \emph{before} the election, but even a signal of the alliances \emph{after} the elections. 
Indeed, our studies show the proximity of the electorate of the governing parties during the electoral debate, beside belonging to different pre-elections alliances.

The paper is structured as follows: 
in Section \ref{results} we describe the performed analyses with a report of the main findings; we summarize research questions, results and this work's contribution in the Section \ref{discussion}. More detailed information regarding the dataset, the analyses and methods can be found in Section \ref{methods} and in the Supplementary Material file.

\section{Results}
\label{results}
\subsection{Identification of alliances via unverified user behaviours}
Using the Twitter API, we have downloaded a sample of all tweets posted from January 28 to March 19, 2018. The query has been performed only requiring each post to contain at least one of a set of Italian elections-related keywords. For more details about the nature of the dataset, refer to the Supplementary Material file, Section I.

As a first step, we have split the sample of Italian-speaking users in two categories, \textit{verified} and \textit{non-verified} users. Each account can request to be verified by the system: by doing so, Twitter guarantees its authenticity. This procedure is generally applied to those people considered of public interest. Therefore, we expect famous people, politicians, newspapers, TV channels, radio channels etc. to have verified accounts, while the remaining users to belong to the other set. Then, we build the bipartite network of retweets between verified and non-verified users: an edge between two users indicates that one has retweeted the other's content at least once during the available time period. At this step we disregard edges' direction, therefore in principle we do not know who is the author of the post and who is the second one who shares the content. However, as shown in Figure 2 of the Supplementary Material file, the actions of tweeting and sharing contents are mostly performed by verified accounts, while non-verified users mainly retweet already published posts.\\

In order to obtain groups of verified users based on their activity, we project the information contained in this bipartite network on the verified users layer. The result of the classical projection methods is a weighted monopartite network: two users are connected if they share at least one common neighbour on the opposite layer and the edge between them is weighted by the number of non-verified users who have simultaneously retweeted their posts. However, this method often generates a dense projection; thus, several procedures has been proposed in order to establish the significance of the edges in the projected network, discounting different pieces of information. For this application we use as benchmark an entropy-based null model that discounts the information contained in the degree sequence of both layers~\cite{Saracco2016}; in this way we are focusing on overlaps that cannot be explained just by the activity of the users. A brief description of this approach can be found in the Section~\ref{methods}, but refer to the Supplementary Material for the details of the entire procedure.\\

Given the projected and validated network of retweets, we have performed a reshuffled community detection procedure, i.e.~the Louvain algorithm~\cite{Blondel2008a} runs several times with a rearranged nodes' ordering and the partition with the highest modularity is selected; in this way we overcome the original algorithm's order dependence~\cite{Fortunato2010}. With this procedure we identify ten groups of non-isolated nodes. However, for the following analyses we focus on a subset of four of them, being those with a remarkable number of nodes (more than a hundred) and a non-trivial interpretation. Two blocks identify quite well the groups of Movimento 5 Stelle (from now on M5S) and right-leaning politicians. For instance, in the former we find the accounts M5S Camera, M5S Senato, M5S Europa and Movimento 5 Stelle, as well as the politicians Danilo Toninelli or Luigi Di Maio. Instead in the latter we see the accounts Forza Italia, Lega - Salvini Premier, Gruppo FI Camera, Fratelli d'Italia, Noi con Salvini as well as the users Silvio Berlusconi, Matteo Salvini, Renato Brunetta or Giorgia Meloni. The two remaining communities are instead more heterogeneous. In one of them, we find a high number of radios, newspapers or newscasts, such as Rai Radio 2, Radio 105, RTL 102.5, Tg Rai, Tg La7, Sky Tg 24, Rai News, la Repubblica, Il Corriere della Sera, Il Post. Since the majority of nodes refers to official accounts of news media, we characterize this group as the one collecting news spreaders and information channels. Finally, the last community encompasses some politicians within the left-leaning parties, such as Matteo Renzi and some other figures belonging to the Democratic party (included the account of the Partito Democratico itself). Therefore we use this interpretation for the last detected community.

\begin{figure}[t]
    \centering
    \includegraphics[scale = 0.30]{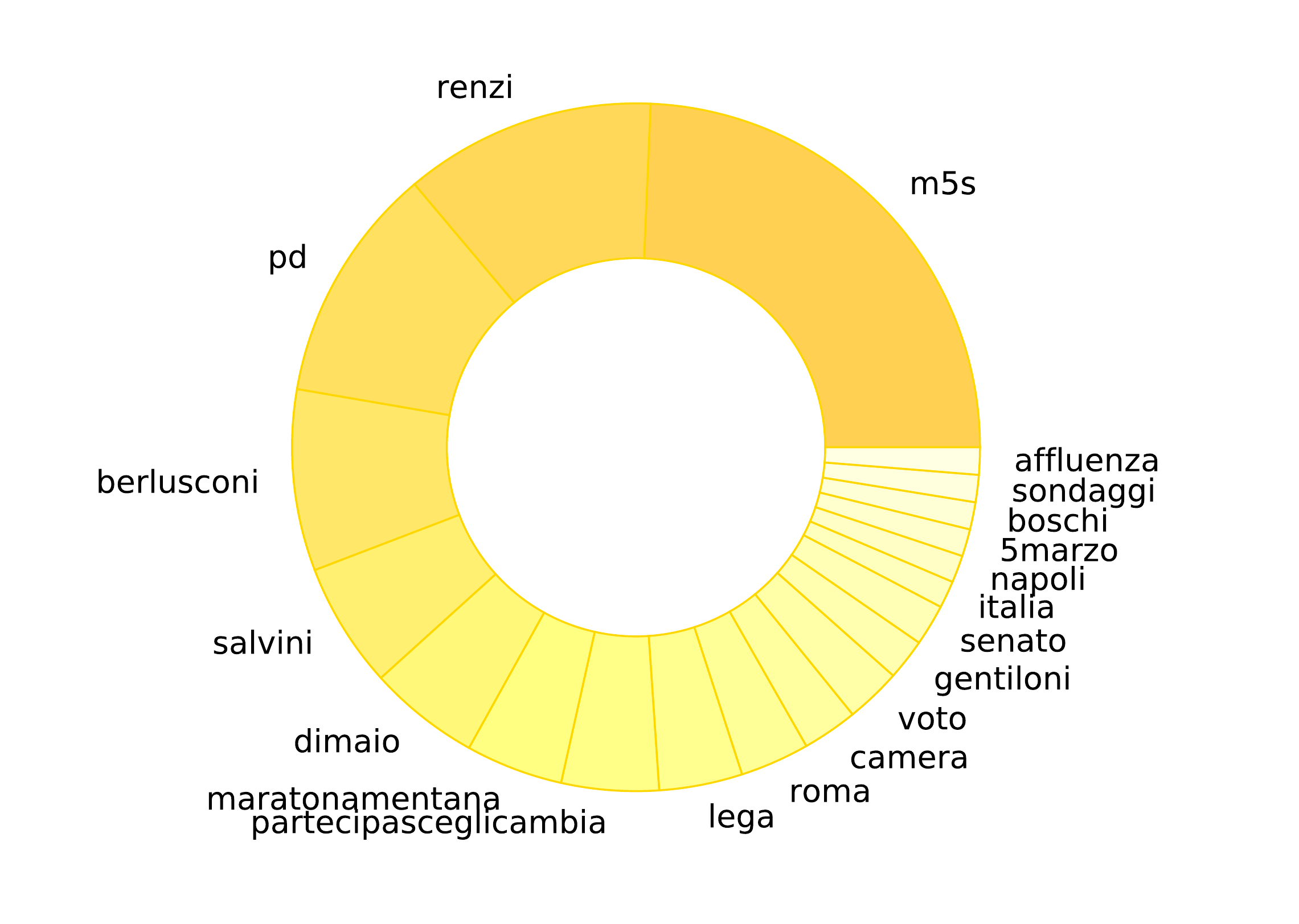}
    \includegraphics[scale = 0.30]{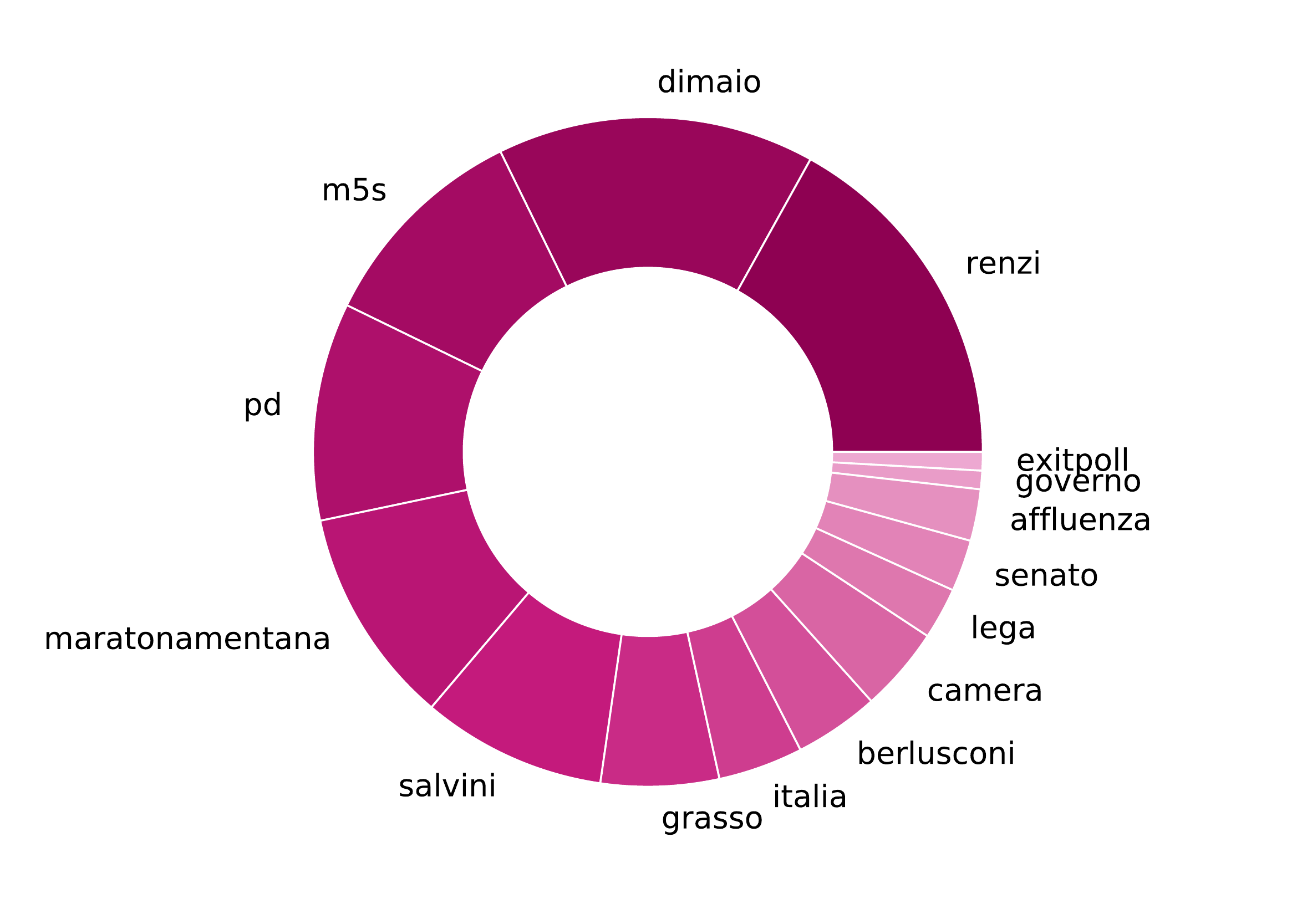}\\
    \includegraphics[scale = 0.30]{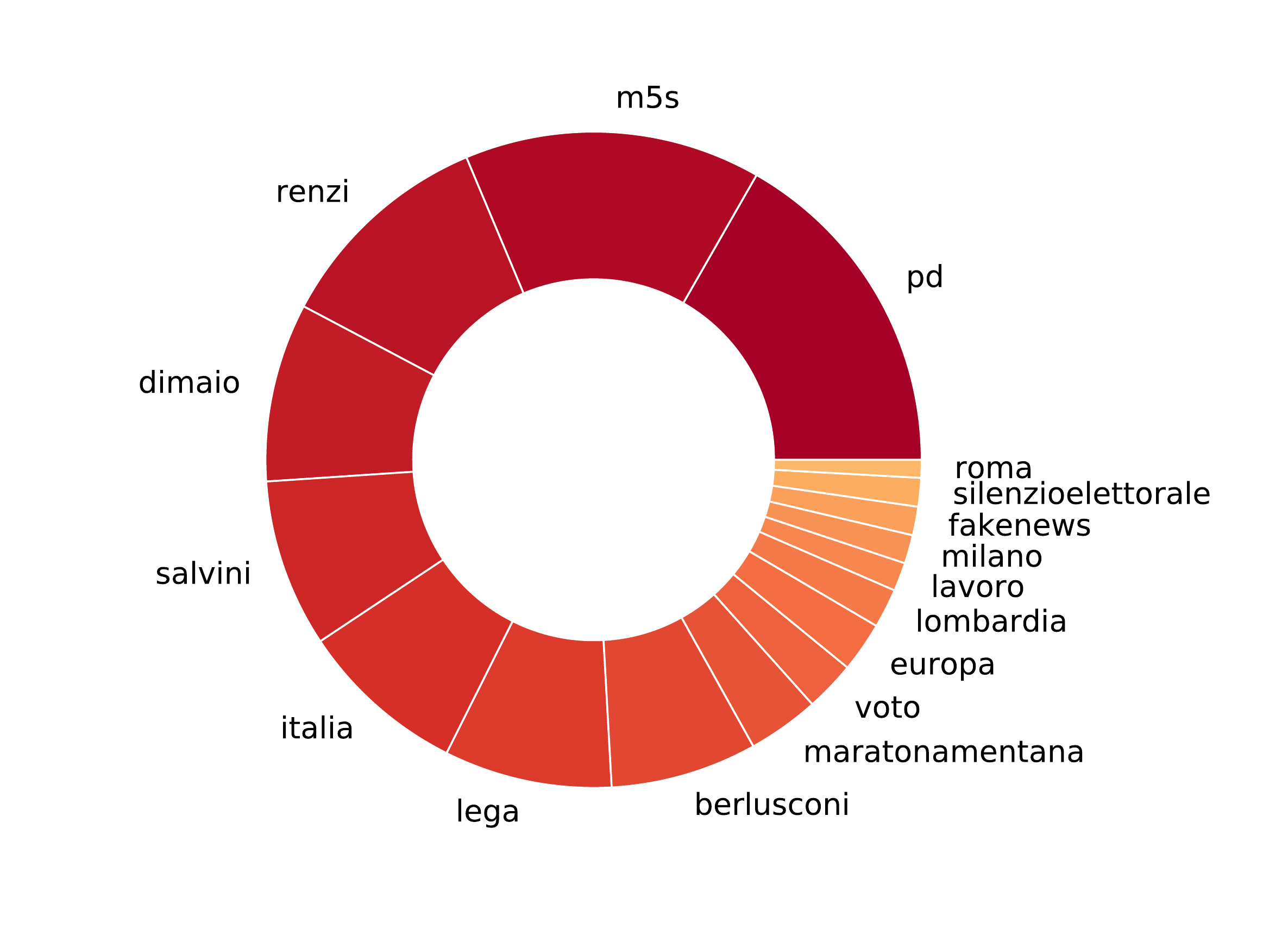}
    \includegraphics[scale = 0.30]{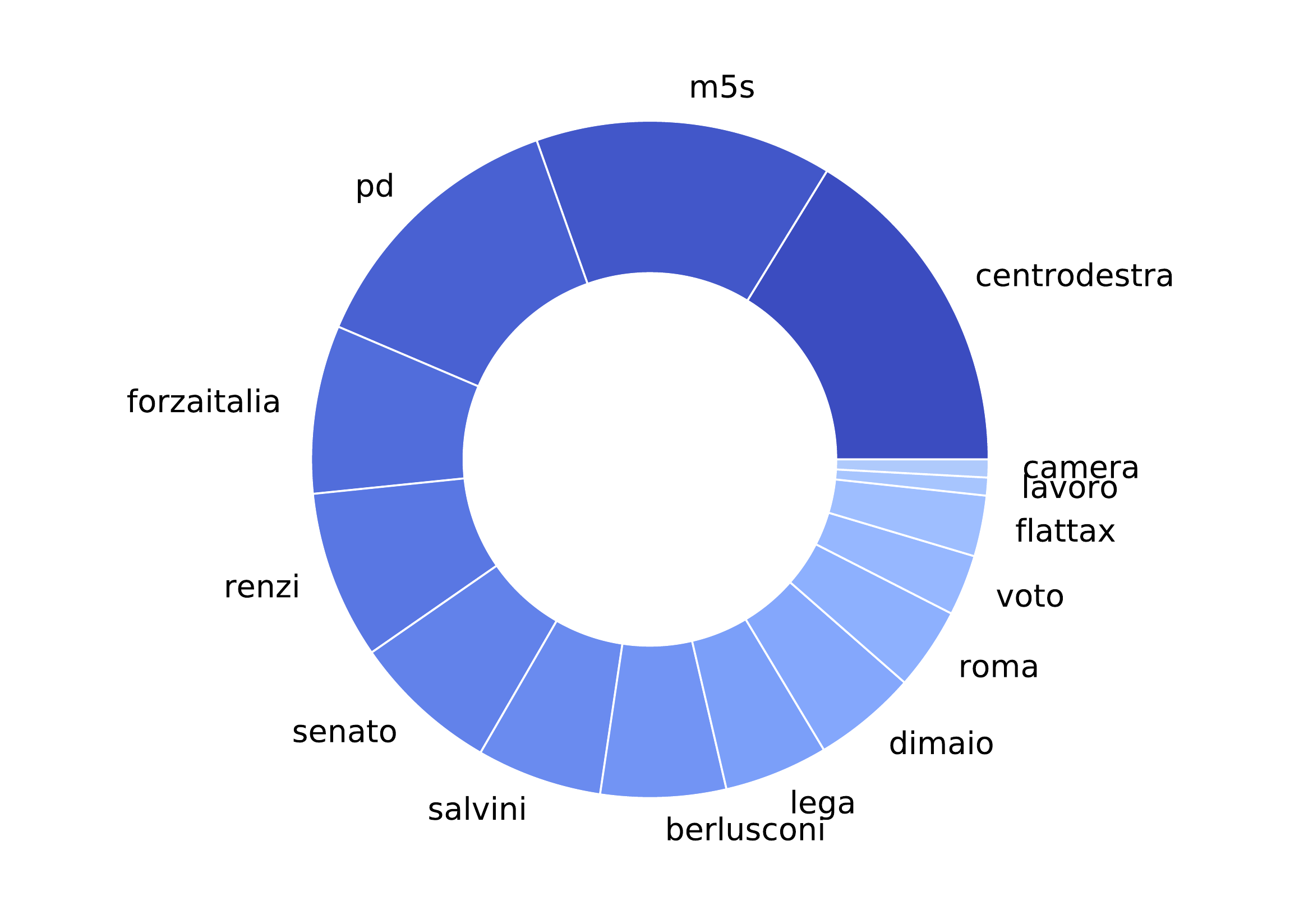}
    \caption{Hashtags with frequence higher than 0.5\% of total number of hashtags used by each group of verified users, excluding ``elezioni'',``elezioni2018'',``elezionipolitiche'',``4marzo'',``4marzo2018'',``marzo2018'',``politiche'', ``politiche2018'', ``elezionipolitiche2018'',``elezioni4marzo2018''. The represented color scales have been obtained rescaling the original frequencies between 0 and 1. }
    \label{fig:pie_plots}
\end{figure}

\subsection{Structure of the communities}
\label{community_structure}
Given the division in political alliances, we start analyzing the topological characteristics of the subgraphs made by each group. An important insight on users' behaviour comes from the observation of the hashtags used by verified users. Excluding the set of keywords used to extract the data, Figure \ref{fig:pie_plots} reproduces the hashtags used more frequently by the verified users of each community, selecting those with a frequency higher than or equal to the 0.5\% of the total number of hashtags in the single community. 
All political (i.e. yellow, red and blue) communities have the name of their own party as the most mentioned tag; indeed we observe, respectively, the hashtags ``m5s'', ``pd" and ``centrodestra'' in their first position. Nevertheless, the second most used  hashtag refers to the main opponent of the political alliance represented by the community. It is curious that this word is the Movimento 5 Stelle for both the right- and left-leaning alliances, since it was effectively the most voted party at the elections. Instead, the second most used hashtag by Movimento 5 Stelle is ``renzi", leader of the Partito Democratico at the moment of the elections, governing at that time.
Also the major exponents of the other parties are mentioned, for instance ``berlusconi'', ``salvini'' and ``dimaio'' appear in the left- and right-leaning parties, as well as in the M5S one and the mixed group. Even if the frequencies of the hashtags are comparable, summarising, all alliances have the name of their own coalition as the most frequent one, followed by the names of the major competitors. More comments about the frequency of hashtags and an analysis of  the information channels consulted by users in the different communities can be found in the Supplementary Materials, Section II.

\begin{figure}[t]
\centering
\includegraphics[scale = 0.61]{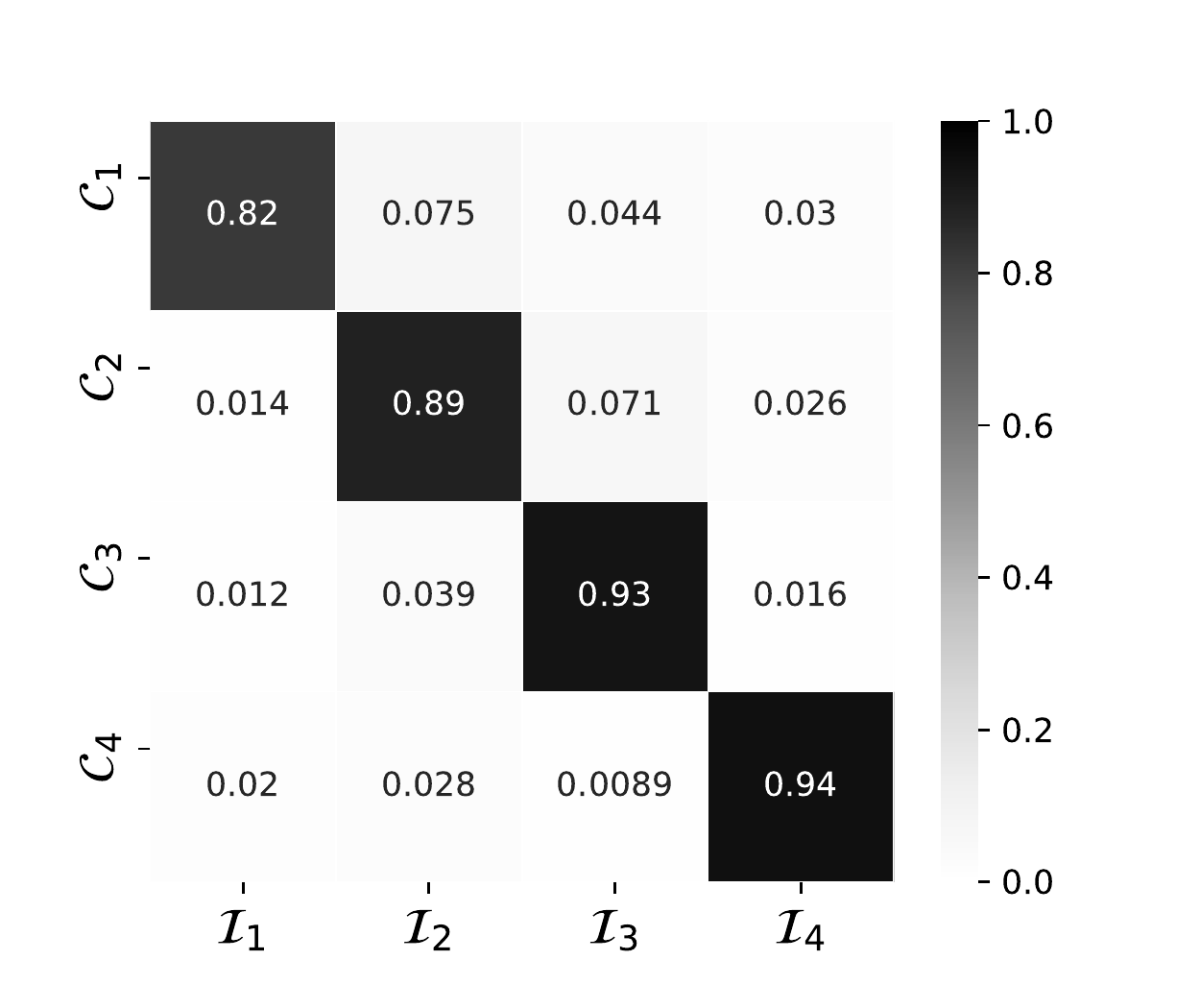}
\includegraphics[scale = 0.43]{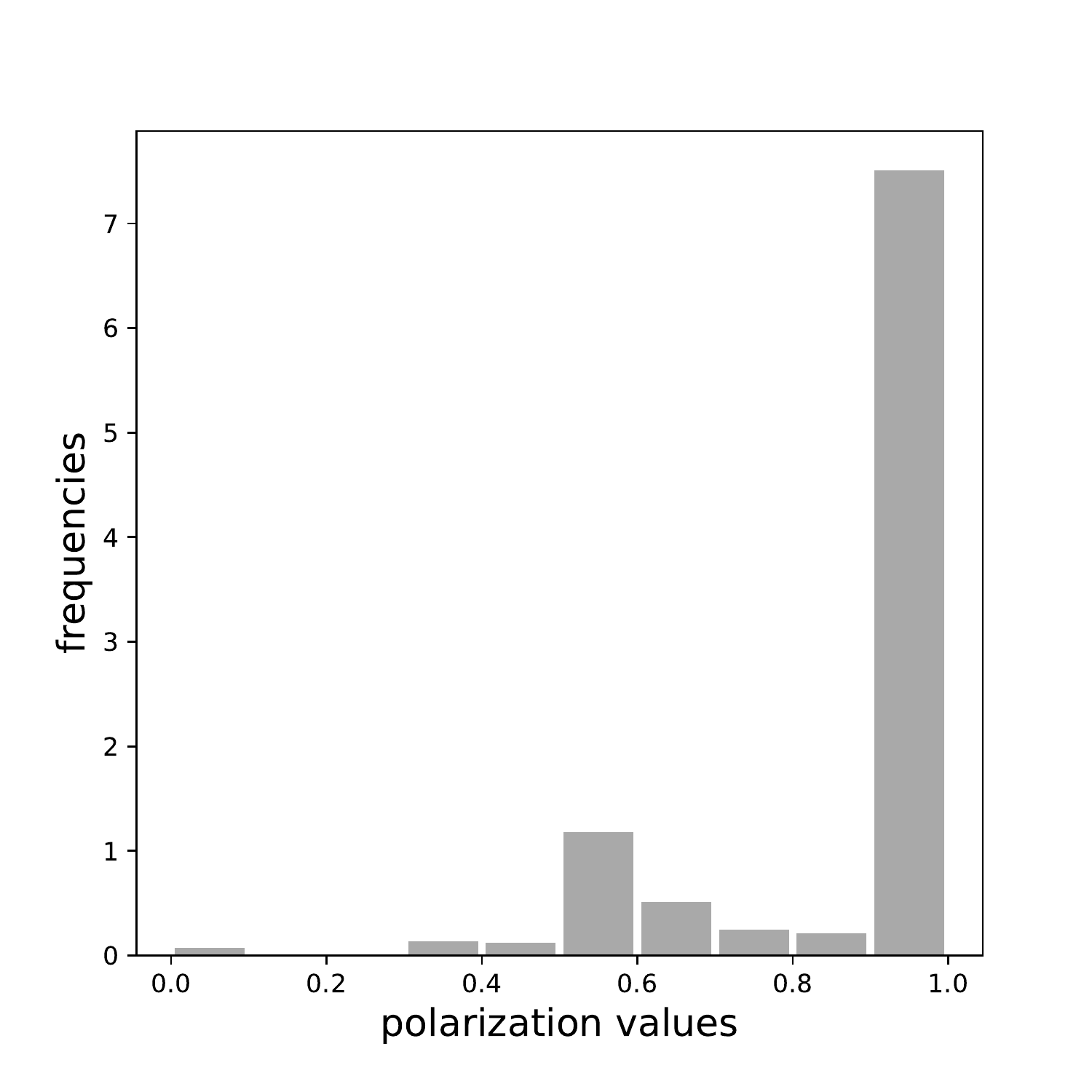}
\caption{Left: heatmap reproducing the average fraction of interactions towards each community, for subsets of non-verified users. Each square in the heatmap indicates the average value of the quantity reported on the $x$-axis, computed over the set of non-verified users belonging to the community on the $y$-axis. Right: histogram of obtained polarization values.}
\label{fig:plot_polarization}
\end{figure}

\subsection{Polarization analysis}
\label{polarisation_section}
Once identified four groups of political characters and information channels  representing the Italian political scenario, we want to analyse how the remaining accounts interact with them. In other words, the goal is to study whether the audience is polarized towards the source of information that better resembles their ideology or uniformly shares contents from accounts of different political orientation.\\
Indicate with $\mathcal{C}_c$ for $c = 1, 2, 3, 4$ the sets of verified users identified at the end of the phase above, denoting respectively the M5S, information channels community, the left-leaning and right-leaning accounts. Also indicate with 
$$\mathcal{N}_\alpha = \{i : i \in L \hbox{ and } m_{i\alpha}^* = 1\}$$ 
the set of neighbours of the non-verified user $\alpha \in \Gamma$ in the bipartite network of verified/unverified users, i.e.~the set of verified users that node $\alpha$ has interacted with. The polarisation index for $\alpha$ is 
\begin{equation}\label{eq:rho}
\rho_\alpha = \max(\{\mathcal{I}_{\alpha,c} : c = 1, 2, 3, 4\}) \hbox{ for } \alpha \in \Gamma
\end{equation}
with 
\begin{equation}\label{eq:i_call}
\mathcal{I}_{\alpha,c} = \frac{|\mathcal{C}_c \cap \mathcal{N}_\alpha|}{|\mathcal{N}_\alpha|} \hbox{ for } c = 1, 2, 3, 4.
\end{equation}
The term $\mathcal{I}_{\alpha,c}$ denotes the fraction of $\alpha$'s interactions towards community $c$, i.e.~the ratio of $\alpha$'s neighbours belonging to community $c$. 
This index has the following characteristics: is bounded in $[0,1]$, therefore $\rho_\alpha = 0$ means that no interaction has been observed with the four groups;
values of $\rho_\alpha$ close to $1/4$ indicate that user $\alpha$ equally interacts with the four clusters; for all the other values, the greater $\rho_\alpha$ the higher the inequality in the number of interactions with the four communities (i.e.~$\rho_\alpha$ close to 1 means that user $\alpha$ almost always has interacted with the same group). \\
\indent The choice of this index has been mostly driven by the observation of users' interactions with the communities. See left panel of Figure \ref{fig:plot_polarization}. Each square in the heatmap reproduces the average value of the quantity on the $x$-axis, computed over the set of non-verified users belonging to the community on the $y$-axis. Most of the non-verified users have an extremely unbalanced distribution of interactions with the members of the political alliances. The pictorial representation of the entire distribution of the terms in \eqref{eq:i_call} within each community is provided in Figure 4 of the Supplementary Material.
Instead, on the right-hand side of Figure \ref{fig:plot_polarization} we represent the histogram of the polarization values obtained for the non-verified users. As introduced by a preliminary evaluation of users' behaviour, many non-verified users show high values of polarization index, meaning that their attention patterns are mostly focused towards a limited group of political characters. 

\begin{figure}[t]
\centering
\includegraphics[scale=0.6]{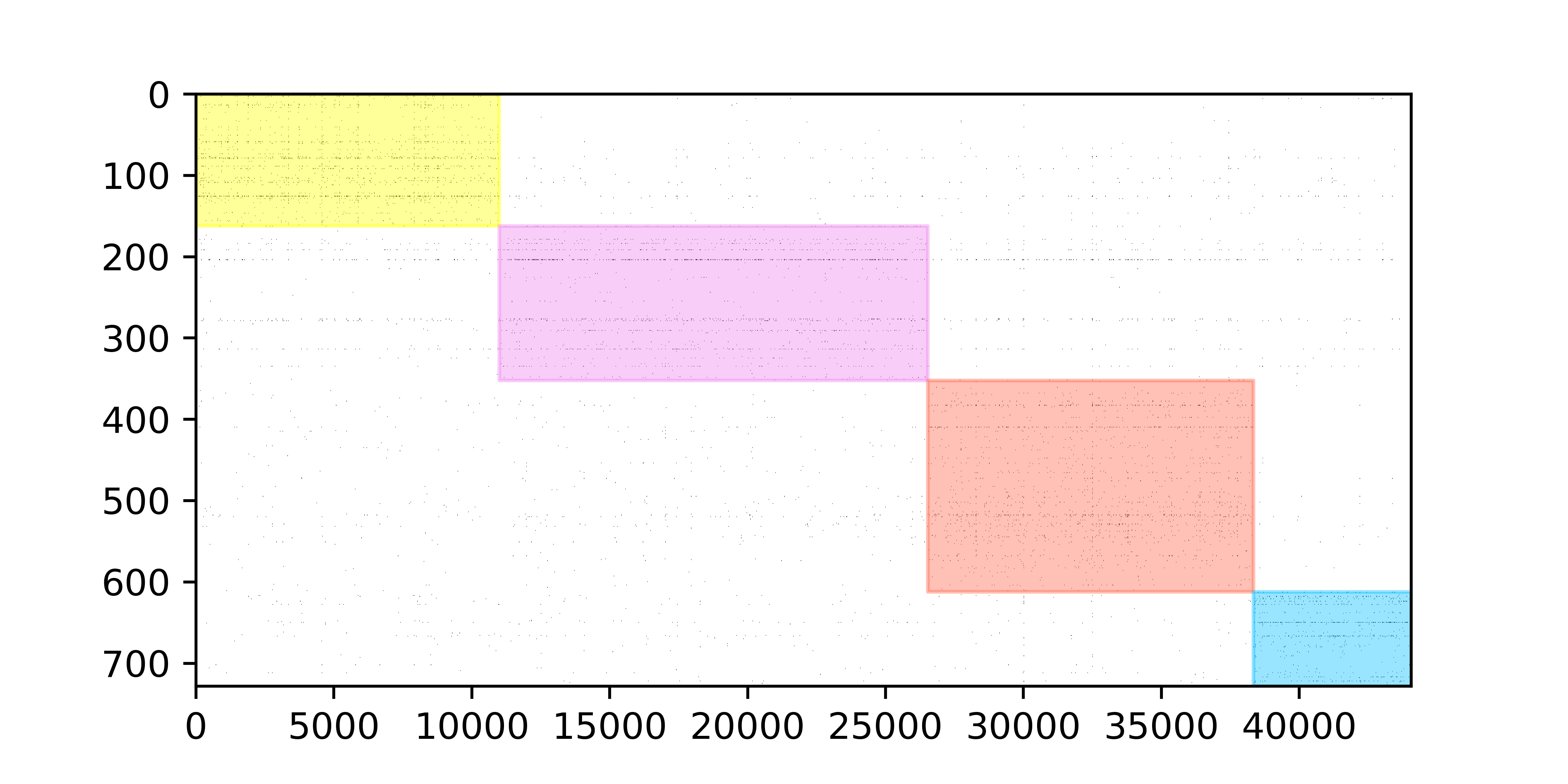}
\caption{Biadjacency matrix of the network of retweets between verified and non-verified users. The coloured blocks identify the four communities of verified users obtained with the community detection method: yellow for M5S, red and blue respectively for left- and right-leaning alliances and purple for the information channels community. The rows of the matrix have been ordered according to the division in communities, while the columns are sorted according to the political affiliation of non-verified users, i.e.~the number of interactions with each group.}
\label{fig:block_matrix}
\end{figure}

Figure \ref{fig:block_matrix} shows the biadjacency matrix of the bipartite network of verified and non-verified users. The coloured blocks identify the four communities obtained with the community detection method: the red and blue blocks respectively identify the groups of left-leaning and right-leaning politicians; the yellow community collects the available political figures within the M5S party, while the violet group represents the information channels community. The rows of the matrix have been ordered according to the division in communities of the verified users, while the non-verified users have been sorted according to their political affiliation, i.e.~the number of interactions towards each group, that is equivalent to the computation of the numerator of the term in equation \eqref{eq:i_call} for all the available communities. Such a matrix exhibits a block structure along the diagonal, indicating a greater number of interactions towards the ``preferred'' community with respect to the others and therefore a higher density of links within the blocks with respect to the external density. 
Contrarily to the majority of studies on the same subject, the polarisation of non-certified users has not been studied simply labeling the units in the verified accounts layer. On the contrary, users in the same community share a significantly high number of non-verified users that have retweeted their content. Indeed, a validated edge between users $i$ and $j$ indicates that a high number of non-verified accounts who shared contents posted by $i$ has also retweeted posts published by $j$ and therefore they are considered similar by a majority of their audience and followers. Therefore, the behaviour of the non-certified accounts is the driver to understand the division in clusters.

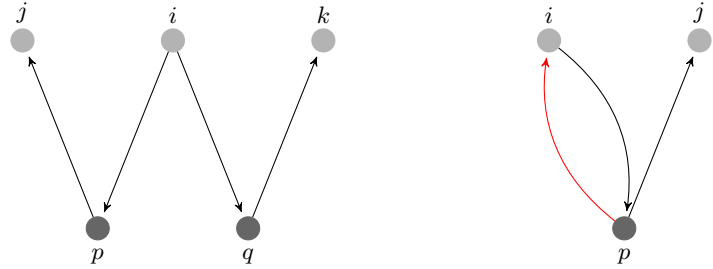
\begin{figure}[t]
\centering
\begin{tikzpicture}[>=stealth',shorten >=2pt,auto,node distance=3cm,
					main node/.style={circle,draw}]
					
\node[label={\small $j$}] (1) at (1, 0) {};
\node[label={\small $i$}] (2) at (3, 0) {};
\node[label={\small $k$}] (3) at (5, 0) {};
\node[label={\small $p$}] (4) at (2, -3.2) {};
\node[label={\small $q$}] (5) at (4, -3.2) {};
\node[main node,fill=black!30,color=black!30] (1) at (1, 0) {};
\node[main node,fill=black!30,color=black!30] (2) at (3, 0) {}; 
\node[main node,fill=black!30,color=black!30] (3) at (5, 0) {}; 
\node[main node,fill=black!60,color=black!60] (4) at (2, -2.5) {}; 
\node[main node,fill=black!60,color=black!60] (5) at (4, -2.5) {}; 
\path[->] (2) edge node {} (5);
\path[->] (2) edge node {} (4);
\path[->] (4) edge node {} (1);
\path[->] (5) edge node {} (3);

\node[label={\small $i$}] (6) at (8, 0) {};
\node[label={\small $j$}] (7) at (10, 0) {};
\node[label={\small $p$}] (8) at (9, -3.2) {};
\node[main node,fill=black!30,color=black!30] (6) at (8, 0) {};
\node[main node,fill=black!30,color=black!30] (7) at (10, 0) {}; 
\node[main node,fill=black!60,color=black!60] (8) at (9, -2.5) {}; 
\path[->] (8) edge node {} (7);
\path[->] (6) edge [bend left] (8);
\path[->,color=red] (8) edge [bend left] (6);

\end{tikzpicture}
\caption{Left: User $i$ is a central user and publishes two posts $p$ and $q$ during the elections. Users $j$ and $k$ respectively retweet the same posts $p$ and $q$ at least one time. Right: illustration of the case in which $i$ retweets one of her/his own posts (the red arrow). This case is excluded from the analysis.}
\label{fig:dir_bip_network}
\end{figure}

\subsection{Influence analysis}
At this point, we proceed with the identification of the significant sources of Twitter viral content. Inspired by the work of~\cite{Bovet2018a}, we propose a \textit{bipartite and directed network of information flow}. An example of this kind of graph is provided in Figure \ref{fig:dir_bip_network}. The users on the upper layer tweet and retweet the posts represented on the lower layer. An out-going edge $(i,p)$ as in the figure indicates that user $i$ has published tweet $p$ during the time period $\mathcal{T}$, while a link in the opposite direction $(p,j)$ denotes that user $j$ has retweeted the same post $p$. In this situation, $j$ behaves as a \textit{spreader}: the larger the number of spreaders of an user, the larger the audience of the contents shared by her/him. Moreover, the larger the number of tweets posted by $i$ that have been retweeted by $j$, the tighter the social bond between the two.\\
As in \cite{Bovet2018a}, we simply project the bipartite and directed network of tweets and retweets onto the users layer: a directed edge between $i$ and $j$ in the projected graph indicates that $j$ has retweeted $i$'s posts at least one time. A list of the nodes with the highest in- and out-degrees per community can be found in the Supplementary Material file, Tables IV and V. However, these findings are not really explanatory in identifying the most \emph{effective} users in the directed network of retweets, since we are missing a benchmark for stating if $j$ is a \textit{significant} contributor of the popularity of $i$'s posts. The identification of such significant tights will be performed in two steps: first, we define a suitable benchmark model to evaluate the significance of pairwise connections; this can be done by using the Bipartite Directed Configuration Model proposed in~\cite{DeJeude2018}, thus discounting the information related to the in- and out- degrees of the nodes of both layers (in the present case the information of in- and out- degrees of posts and users). Then we extend the validation presented in \cite{Saracco2016} to this kind of graphs, in order to identify the significant tights between source and spreader and validate the directed network of information flows. The overall procedure is explained in the Methods section (Section \ref{methods}), but more details can be found in the Supplementary Materials. In the final, validated and directed network of users, only the pairs of users in which one of the two retweets the other's contents more than what is expected by the considered null model will be connected.\\

\begin{figure}[tp]
\includegraphics[scale = 0.37]{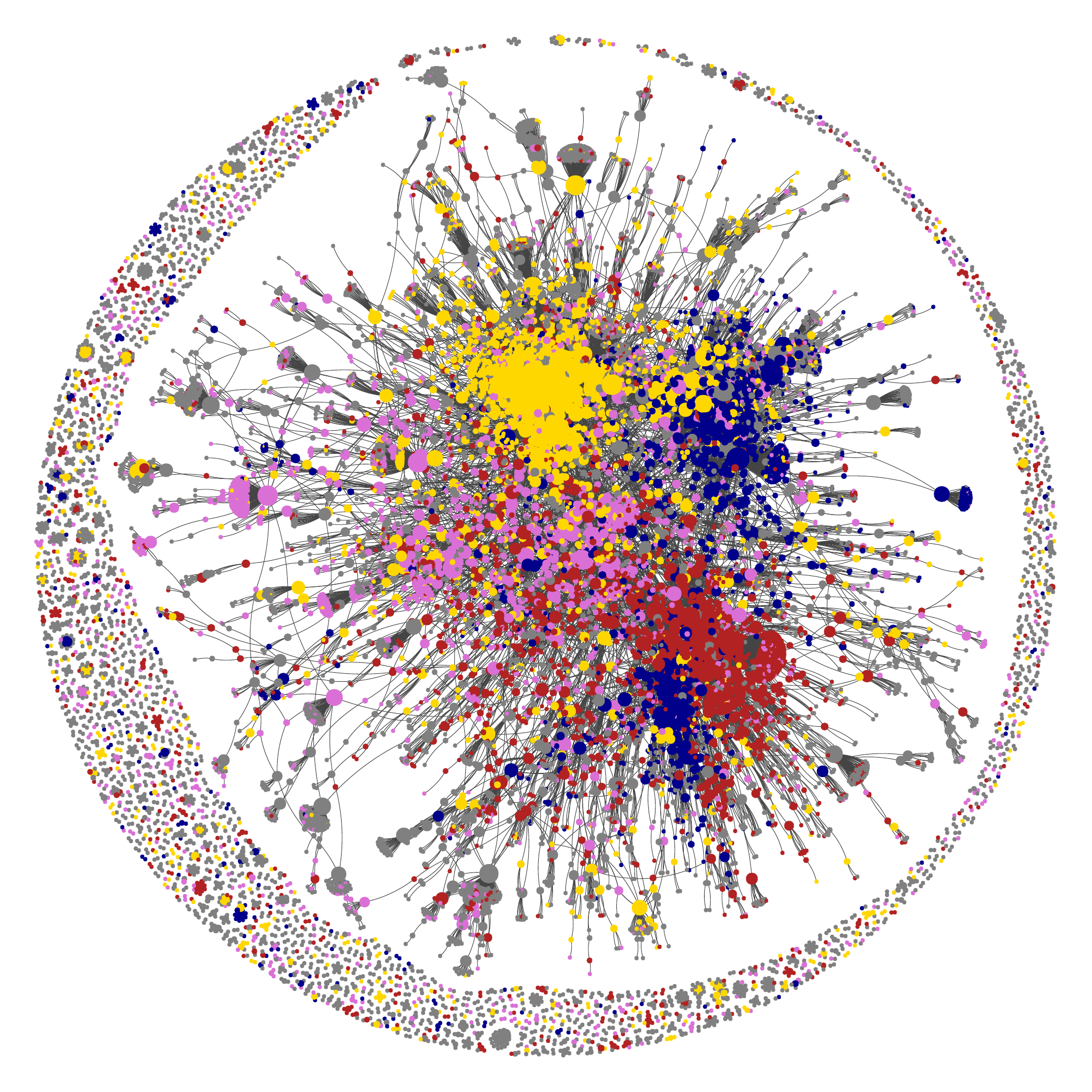}
\caption{Entire directed network of retweets after the bipartite validation procedure. An edge from $i$ to $j$ indicates that $j$ has retweeted $i$ a significantly high number of times. Nodes’ colour identifies the community to which the user belongs (red for left-leaning, yellow for M5S, purple for information channels and blue for right-leaning communities) while nodes’ dimension indicates their out-degree in the validated graph (i.e. how often their posts have been retweeted a significantly high number of times). Note that the blue nodes are divided into 2 subcommunities, one closer to the M5S community, the other closer to the left-leaning one.}
\label{fig:ret_val_net}
\end{figure}

\begin{figure}[h]
    \centering
    \includegraphics[scale = 0.6]{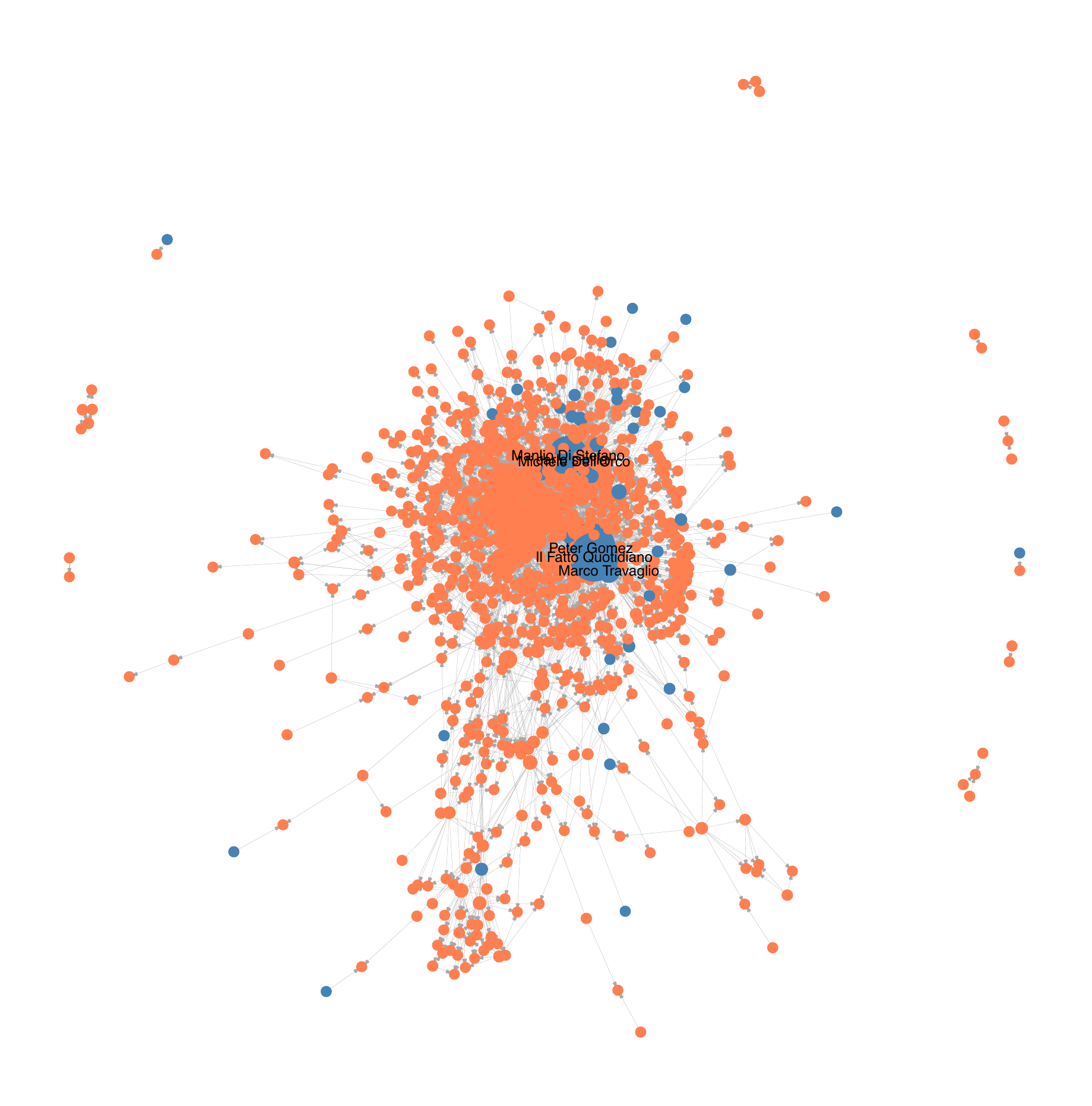}
    \includegraphics[scale = 0.6]{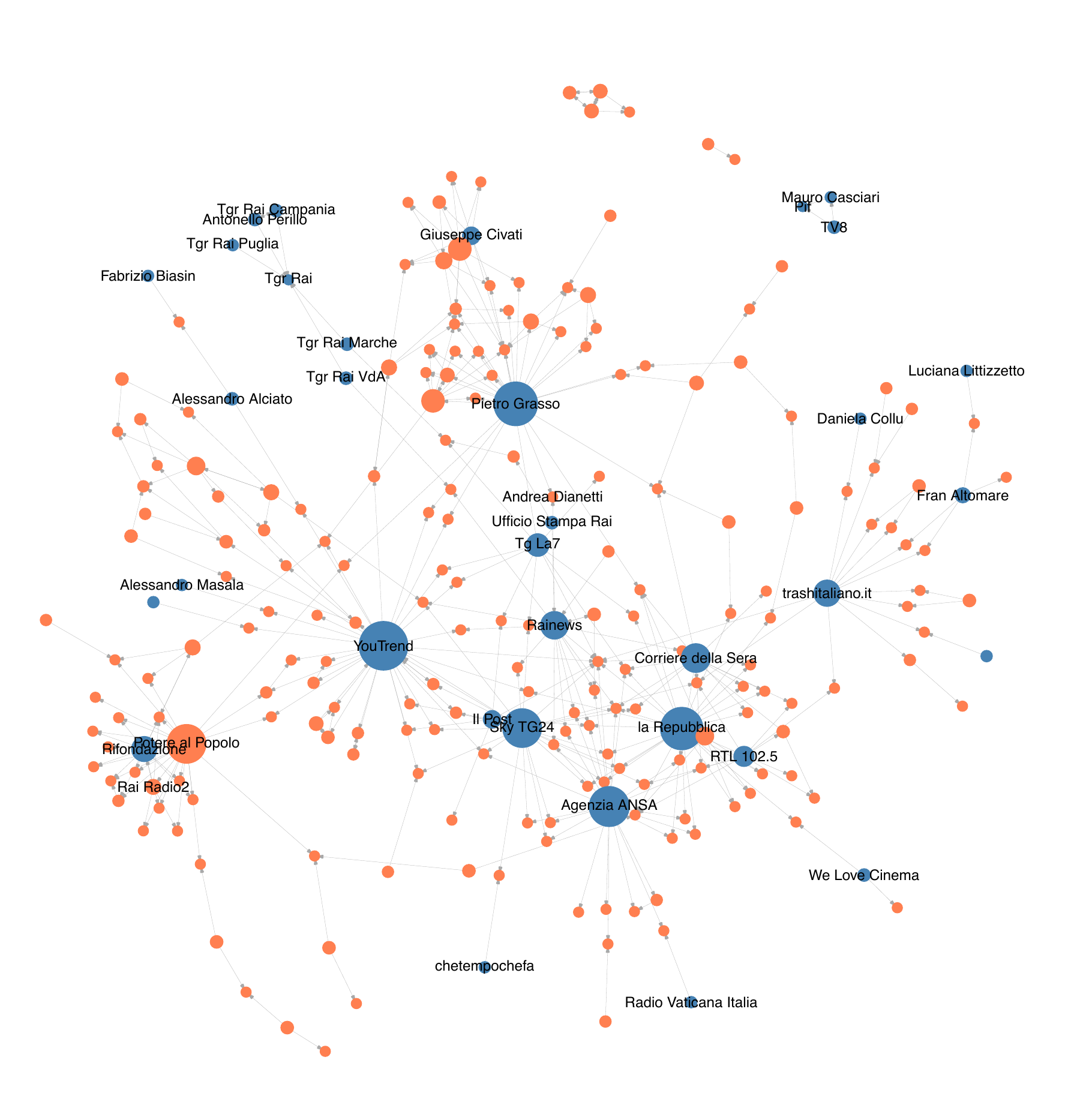}
    \caption{Most retweeted users in groups $\mathcal{C}_1$ and $\mathcal{C}_2$. Nodes' dimension is proportional to the out-degree in the subgraph generated by the community (i.e.~number of retweets received from users from the same community). The blue color indicates that the node is verified, while orange is not verified.}
    \label{fig:out_degrees_1}
\end{figure}

\begin{figure}[h]
    \includegraphics[scale = 0.63]{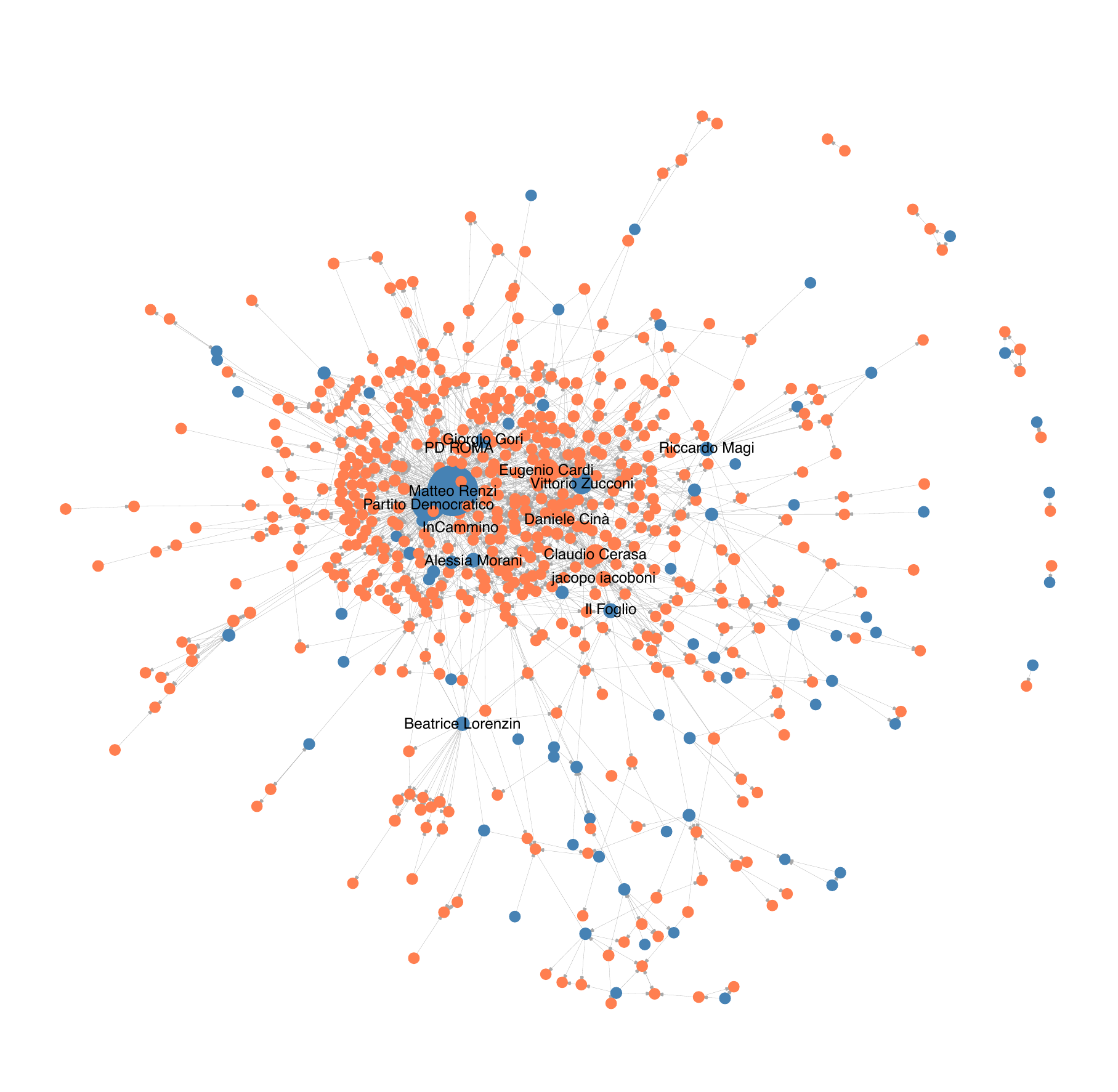}
    \includegraphics[scale = 0.63]{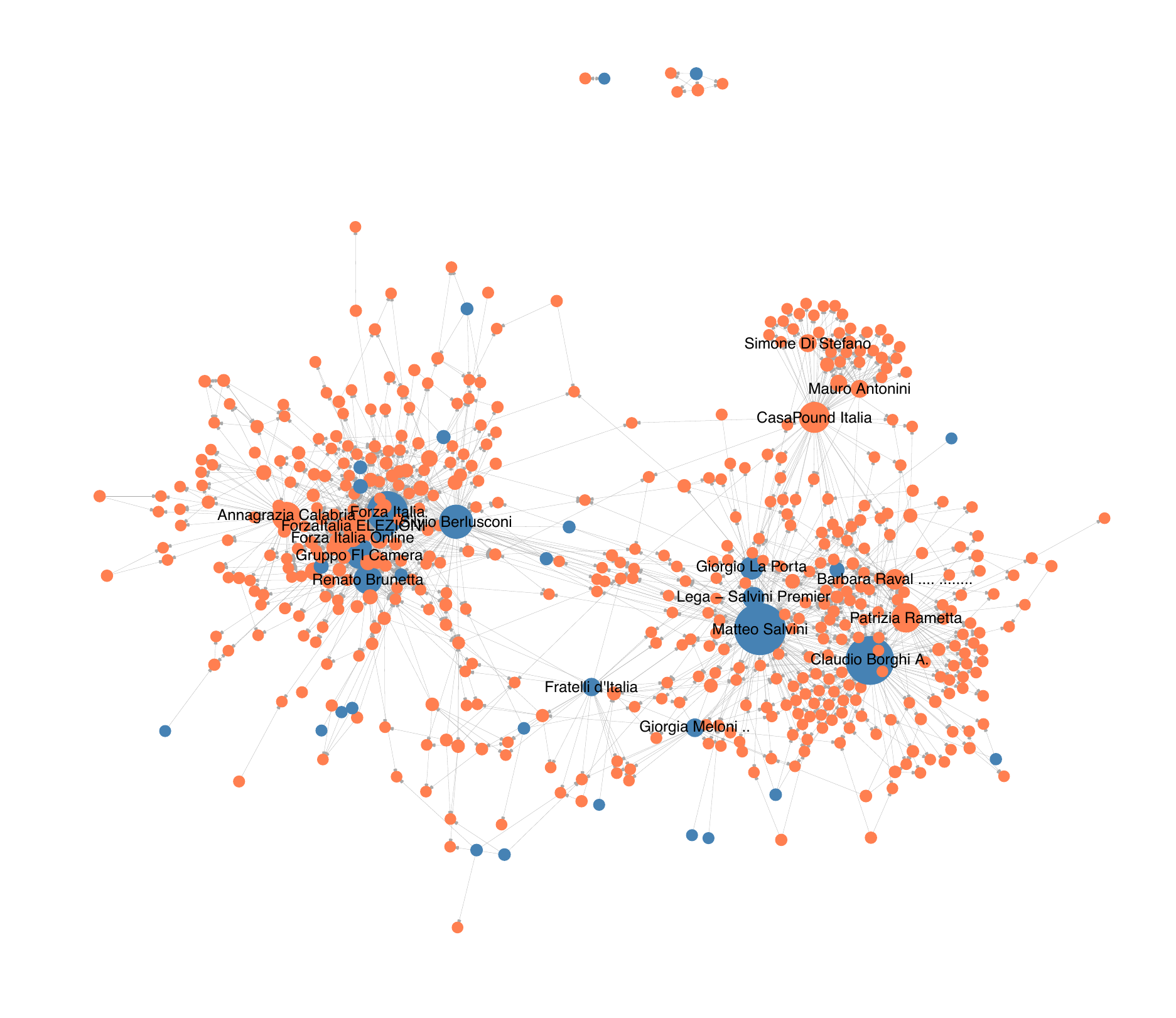}
    \caption{Most retweeted users in groups $\mathcal{C}_3$ and $\mathcal{C}_4$. Nodes' dimension is proportional to the out-degree in the subgraph generated by the community (i.e.~number of retweets received from users from the same community). The blue color indicates that the node is verified, while orange is not verified. }
    \label{fig:out_degrees_2}
\end{figure}

A pictorial representation of the network of validated retweets is provided in Figure \ref{fig:ret_val_net}. Nodes' colour identifies the user's community while nodes' dimension indicates their out-degree in the validated graph. The structure of this network is better represented in Figures \ref{fig:out_degrees_1} and \ref{fig:out_degrees_2}. Each plot focuses on the structure of the subgraphs of the directed network generated by each community. Nodes dimension is directly proportional to their out-degrees in the subgraph, therefore the larger the node, the higher the number of times that user has been retweeted by the other accounts. Nodes colour is instead related to whether the account has been verified or not: blue for verified users, orange for non-verified ones.\\

The first plot is related to the M5S community and shows strongly connected block of (mostly non-verified) nodes that retweets among themselves and with the verified accounts of the community, the most central of which are the Twitter accounts of the newspaper Il Fatto Quotidiano and its journalists Marco Travaglio, Peter Gomez and Antonio Padellaro. In the other communities journalists and newspapers do not form such a strong core. It is interesting to notice that the M5S political leader Luigi Di Maio does not belong to this big community, but is located in a small community outside this large component.

The second plot represents the purple community. The most central nodes are the verified accounts of newspapers (see for example La Repubblica or Il Corriere della Sera) and information channels (such as Sky TG24, Tg La7, Agenzia Ansa or Rainews). Some politicians such as Pietro Grasso or Giuseppe Civati are present in this group, together with the political parties of Rifondazione and Potere al Popolo. These politicians represent the more extreme, left-leaning orientation and that have not encountered a commonality of interests and supporters with the accounts in the community of Partito Democratico and indeed they belong to different communities.

In the plot associated to the left-leaning community (the third one) we identify a central block of mostly non-verified users. The most retweeted figures are Matteo Renzi and the account of Partito Democratico, as shown by their high values of out-degrees. The remaining verified nodes are mostly well-known characters in the political scenario (such as Maria Elena Boschi and Carlo Calenda), as well as newspapers (see for example Il Foglio or IlSole24Ore). Among the non-verified users we have the accounts of the Partito Democratico political parties related to the areas of Milan and Rome.

Finally, the plot associated to the right-leaning community is characterised by two quite separate clusters; one of them is centered on the accounts of people belonging to Lega Nord, such as Matteo Salvini, Claudio Borghi and the party Lega-Salvini Premier. On the other side there are the accounts of Forza Italia and Gruppo FI Camera and some of its exponents like Silvio Berlusconi or Renato Brunetta. The two verified nodes of Giorgia Meloni and Fratelli d'Italia (the political party she is leading) receive retweets from both sides, nevertheless being closer to the Lega pole. Another popular node is CasaPound, that has its own circle of retweeters and share some interactions with the subgroup of Lega Nord.


In order to understand how viral news propagate through the validated network, we analyse the percentage of retweets coming from inside of the same community and the percentage that, instead, derives from the other ones. The results are shown in Figure 1 of the Supplementary Materials and are extensively commented therein.

\section{Discussion}
\label{discussion}
In this work, we focused the attention on the identification of the sources of Twitter viral content during the last Italian electoral campaign of 2018. We gathered the data from the Twitter API during the month before the elections. As a first step, we exploited the way people consume news on the social media to identify four groups of strongly connected verified users: two certified users are connected in the validated network of retweets if a significantly high number of non-certified accounts retweets content published by both of them. By construction, the behaviour of non-certified users is exploited to understand the division of the political sphere into clusters. An analysis of the obtained communities shows that the most central accounts are mostly newspapers, journalists and information channels in general, each of them belonging to one of the previously listed clusters (see Table II of the Supplementary Material file). Looking at the hashtags included in the posts published by the groups, we see that the most used keywords are referred to the party itself and its members, followed by keywords related to political competitors. Given the obtained division in political alliances, we studied the behaviour of the remaining non-verified users towards these groups. More specifically, we have observed the fraction of retweets directed towards each alliance: we observe a strongly polarized behaviour, since the majority of the uncertified accounts in the bipartite network of retweets mostly interact with one community only. In order to strengthen this result, we also perform a different analysis comparing the distribution of polarization values observed for users with the same number of interactions (see Figure 5 of the Supplementary Material). Also in this case, the distribution is skewed towards the higher numbers, indicating a focus towards the same group of users.

As a second step, we focused our attention on the identification of significant news spreaders. Following the methodology presented in \cite{Bovet2018a}, we constructed the directed network of retweets among users and selected the names with the highest out-degree and in-degree. 
In order to statistically validate our findings, we constructed the bipartite and directed network of tweets and retweets introduced in Section \ref{bipartite_directed_cm} and performed the validation procedure described therein. The outcome of this analysis is a monopartite, directed network of users, in which an edge from $i$ to $j$ indicates that the latter retweeted contents posted by the former a significantly high number of times. The visualization of this validated network  helps to understand the actual composition of each coalition, as well as the possible interconnections between them. For instance, we observed that approximately 8\% of the connections between one community and another one happen between verified and unverified users, where, in most of the cases, the latter retweeted some posts from the former. 
However, we also see connections involving newspapers and information channels belonging to different coalitions, confirming again their essential role and centrality in spreading news on the social networks. Finally, we analysed the origin of the retweets received by each community: even though the distribution seems less polarized in this case, a higher percentage of the retweets received by a community comes from users belonging to the community itself, especially for the cases of $\mathcal{C}_3$ and $\mathcal{C}_4$, showing that most of the interactions still comes from the same sphere of influence (see Figures 1 and 6 of the Supplementary Material). 

In our view, the methodological contributions of this paper are manifold. 
First, at odds with a majority of papers dealing with the same topic, our dataset was not manually labelled: we identified groups of strongly connected users starting from the behaviour of non-certified ones and their interactions with certified people. Despite this data-driven approach, we manage to identify four clusters of users that are closely aligned with the known Italian political division. Therefore, how people consume the news and interact with main political figures helps to shed light on the actual division of the verified users according to their political orientation. Our second contribution resides in the representation of the network of activities on Twitter as a bipartite, directed network. We employed the null model for directed bipartite networks proposed in \cite{DeJeude2018} to identify the significant information flows between pairs of users and different communities. We observe that the right-leaning alliance is divided in two subcommunities: one centered on Berlusconi and Forza Italia, closer to the Partito Democratico community; the other, led by Salvini and Lega party, closer to the M5S community. The result is striking since M5S and Lega actually allied to form a new government, after the elections, when no predefined alliance obtained the absolute majority.  Our analysis suggests that seeds of this commonalities of interests were already present in the way different electors acted on Twitter.

Interesting lines of research, left for future investigations, include a validation of our approach to a corpus of different elections, in order to identify potential regularities or differences between countries, and the use of tools from natural language processing to infer how positively- and negatively-connoted tweets are distributed across the communities.

\section{Methods}
\label{methods}

\subsection{Definition of the polarisation network}
\subsubsection{Bipartite network of verified/unverified users}\label{data_description}
As a first step, we have split the sample of available and Italian-speaking users in two categories, the groups of \textit{verified} and \textit{non-verified} users. 
Each account can request to be verified by the system: by doing so, Twitter guarantees that the account is authentic. This kind of procedure is, in general, applied to all those people who are considered of public interest. Therefore, we expect the accounts of famous people, politicians, newspapers, TV channels, radio channels etc. to be included into the set of verified users, while all the remaining users to belong to the other set. 
Given this division, we construct a first bipartite network from the data, the  network of retweets between verified and non-verified users during the whole period $\mathbf{G}^*_\text{Bi} = (\text{L}, \Gamma, \text{E})$\footnote{In the following we will refer to quantities regarding the real network with an asterisk $*$.}.
By definition of bipartite networks, with this representation we exclude all the cases in which two certified (or non-certified) users retweet each other. We do so since we are interested in exploiting the way normal people consume the news to detect connected groups of certified users. Refer to Section V A 1 of the Supplementary Materials for a more detailed description of this network.

\subsubsection{Bipartite Configuration Model and undirected validated projection}
\noindent In order to evaluate the relevance of the bipartite network of retweets between verified and non-verified users, we need to define a suitable  model as a benchmark for our measurements. Indeed, at the present step we want to discount the information due to the activity of the users, both verified and unverified, in order to detect non-trivial superposition of connections. We thus implement a Bipartite Configuration Model (BiCM)~\cite{Saracco2015a} for this case, that maximise the entropy of the system, constraining the degree sequence of the two layers. A full detailed description of the null model can be found in the Supplementary Materials. Let us just remind here that at the end of this procedure we obtain a probability per graph that factorises in independent probabilities per link.\\ 
In order to infer information about similar behaviours of verified users, we first project the information contained in the bipartite network $\mathbf{G}^*_\text{Bi}$ on the verified-users layer. 
The result of the classical projection methods is a weighted monopartite network: two users $i,j \in L$ are connected if they share at least one common neighbour on the opposite layer $\Gamma$ and the edge between them 
is weighted by the number of non-verified users who have simultaneously retweeted their posts. This quantity is expressed by the number of $V-$motifs between users $i$ and $j$~\cite{Diestel2012, Saracco2015a} 
$$V_{ij}^* = \sum_{\alpha \in \Gamma} m_{i\alpha}^*m_{j\alpha}^*.$$ 
However, this method often generates a highly connected projection; in order to extract the statistically significant information of this projection we implement the method of~\cite{Saracco2016}. Roughly speaking, the method consists in comparing the realised $V-$motifs with the expectation of the Bipartite Configuration Model. In this sense, the statically significant $V-$motifs are extracted: otherwise stated, all motifs that cannot be explained by the degree sequence only are validated. More details can be found in the Supplementary Materials, Sections V A 2 - V A 3.

\subsection{Directed network of information flow}
\subsubsection{Bipartite directed network of tweetin/retweeting activity}
In this section, we describe the construction of the bipartite and directed network of information flow. 
A directed edge $(i,p)$ indicates that user $i$ is the original creator of post $p$, while an edge on the opposite direction $(p, j)$ shows that $j$ has retweeted tweet $p$ \textit{at least one time} during the elections days. Such a situation is shown in the left panel of  Figure \ref{fig:dir_bip_network}. In the picture, $i$ is a user who publishes two posts $p$ and $q$ during the election days; the users $j$ retweets post $p$ and user $k$ retweets post $q$. Notice that, by construction, we cannot keep track of the chains of sequential retweets (consult Section I of the Supplementary Materials for more details about the nature of the data set). Therefore, the edge $(q,k)$ does not necessarily mean that $k$ has retweeted directly from $i$'s post: it is possible (though less likely) that $k$ has retweeted post $q$ from one of user $i$'s retweets. Another particular case is self-retweet: Twitter allows the users to retweet their own posts, either directly from the original tweet or from somebody else's retweet. With this network representation, these cases can be illustrated as in right panel of Figure \ref{fig:dir_bip_network}, where the retweet by $i$ itself is indicated with the red arrow. However, these type of edges have been excluded from the analysis, since they represent a very small percentage of the overall number of retweets observed in the data (approximately 0.97\%) and we are rather interested in how people from the same political coalition interact with each other to boost the visibility of their opinion. \\
\indent This bipartite and directed network $\mathbf{G}^*_\text{BiD}$ can be represented by means of two biadjacency matrices, one for the tweets $\mathbf{T}^*= \{t_{ip}^* : i \in U \hbox{ and } p \in P \}$ and the other one for the retweets $\mathbf{R}^* = \{r_{pj}^* : p \in P \hbox{ and } j \in U\}$. Essentially $t_{ip}^* = 1$ if user $i$ has tweeted post $p$ and 0 otherwise and $r_{pj}^* = 1$ whether or not at least one retweet to post $p$ by user $j$ is observed during $\mathcal{T}$. See Section V B for a more detailed description of this directed network.

\subsubsection{Bipartite Directed Configuration Model and directed validated projectiion}
\label{bipartite_directed_cm}
For the actual analysis we are interested in determining which are the most pervasive tweet flows, discounting both the activity of the users and the virality of the tweets. Otherwise stated, we want to measure which are the non-trivial patterns observed in the network, not explained by the degree sequence for both layers and highlight non-trivial patterns describing the structure of the actual system. The randomization of a bipartite directed network with constraints on the in- and out-degree sequences was first presented in~\cite{DeJeude2018}; we revise the procedure that leads to the definition of the probabilities in the Supplementary Materials. Let us remark that, as in the previous (undirected) case of the BiCM, the probability per graph factorises in terms of probabilities per (directed) link. Moreover, respect to the previous case there is a crucial simplification: indeed each post cannot have more than an author, thus the in-degree of tweet posts is $\langle\kappa_p^\text{in}\rangle=1$ over the ensemble defined by Bipartite Directed Configuration Model (BiDCM).\\
Extending the definition of $V-$motifs to directed networks, the quantity $\mathcal{V}_{ij}$  identifies the number of times $j$ has acted as a spreader for $i$, retweeting content posted by her/him~\footnote{Note that $\mathcal{V}_{ij}\neq\mathcal{V}_{ji}$.}. Using the null model's expectation of this quantity we are able to infer the statistical significance of the tight between $i$ and $j$, following a procedure similar to the one of~\cite{Saracco2016}; consult the Supplementary Materials for further details. This procedure returns a squared $N_U \times N_U$ matrix collecting the significance of each link and each edge is treated as a separate null hypothesis to test~\footnote{Actually, there is a tricky part. Since all users that were not author of retweeted posts have probability exactly equal to zero to spread a tweet, their probability to be the source of a directed $V-$ motif is again exactly zero, for each of the possible tweet. Those cases were not considered in the FDR.}. Only the links associated to rejected hypotheses have to be included in the validated network: whenever $\mathcal{V}_{ij}^*$ is statistically validated, $j$ is considered a significant spreader of $i$'s tweets and the directed edge $(i,j)$ is included in the validated network of information flows $\mathbf{F} = (f_{ij})_{ij \in U}$. Refer to Sections V C - V D for a more detailed description of these methods.

\section*{Acknowledgements}
This work of F.S. and G.C. was supported by the EU projects CoeGSS (Grant No. 676547), Openmaker (Grant No. 687941), SoBigData (Grant No. 654024). G.C. is supported also by EC TENDER SMART (Grant No. 2017/0090 LC-00704693).

\section*{Author Contributions}
F.S. collected the data. C.B and F.S. designed and performed the analyses and prepared tables and figures. All authors wrote, reviewed and approved the manuscript.

\section*{Additional Information}
The authors declare no competing interests.

\section*{Data availability}
The datasets generated and analysed during the current study are not publicly available due to the data cleaning procedure the authors implement for research purposes, but are available from the corresponding author on reasonable request.

\bibliographystyle{unsrtnat}
\bibliography{references}

\end{document}